\documentstyle[rotate,epsfig]{lamuphys}
\def\Journal#1#2#3#4{{#1} {\bf #2}, #3 (#4)}

\def\NPA{{\em Nucl. Phys.} A}
\def\NPB{{\em Nucl. Phys.} B}
\def\NPP{\em Nucl. Part. Phys.}
\def\PLB{{\em Phys. Lett.} B}

\def\SJNP{\em Sov. Jour. Nucl. Phys.}
\def\PRL{\em Phys. Rev. Lett.}

\def\PRD{{\em Phys. Rev.} D}
\def\ZPC{{\em Z. Phys.} C}

\makeatletter
\let\chapter\hid@chapter
\makeatother

\begin{document}
\pagenumbering{arabic}

\title{Pion and Kaon Polarizabilities\\ and Radiative Transitions}

\author{Murray A. Moinester and Victor Steiner}

\institute {School of Physics and Astronomy, 
R. and B. Sackler Faculty of Exact Sciences,\\
Tel Aviv University, 69978 Ramat Aviv, Israel,\\
e-mail: murraym$@$silly.tau.ac.il, steiner$@$gluon.tau.ac.il}

\maketitle

\leftline{\it Tel Aviv U. Preprint TAUP-2469-97,} 
\leftline{\it Contribution to the Workshop on Chiral Dynamics, U. Mainz, Sept. 1997,} 
\leftline{\it Springer-Verlag 1998, Eds. A. Bernstein and Th. Walcher}

\begin{abstract}
CERN COMPASS plans measurements of $\gamma\pi$ and $\gamma$K interactions using
50-280 GeV pion (kaon) beams and a virtual photon target. Pion (kaon)
polarizabilities and radiative transitions will be measured via Primakoff effect
reactions such as $\pi^- \gamma \rightarrow {\pi^-}' \gamma$ and $\pi^- \gamma
\rightarrow meson$. The former can test a precise prediction of chiral symmetry;
the latter for $\pi \gamma \rightarrow a_1(1260)$ is important for understanding
the polarizability. The radiative transition of a pion to a low mass two-pion
system, $\pi^- \gamma \rightarrow {\pi^-}' \pi^0$, can also be studied to measure
the chiral anomaly amplitude F$_{3\pi}$ (characterizing $\gamma \rightarrow  3
\pi$) arising from the effective Chiral Lagrangian. We review here the motivation
for the above physics program. We describe the beam, target, detector, and
trigger requirements for these experiments. We also describe FNAL SELEX 
attempts to  study related physics via the interaction of 600 GeV pions with
target electrons. Data analysis in progress aims to identify the
reactions $ \pi e \rightarrow \pi' e'  \pi^0$ related to the chiral anomaly, and
$\pi e \rightarrow \pi' e' \gamma$ related to pion polarizabilities.
\end{abstract}

\section{Introduction} 

Pion and kaon polarizabilities and associated radiative transitions may be measured
in the CERN COMPASS experiment (\cite{paul}, \cite{cd}). Hadron ($h$) radiative
transitions may be measured by the Coulomb coherent production reactions $ \gamma h
\rightarrow h{^*}$, for pions and kaons. The cross sections of these reactions are
proportional to the radiative widths $\Gamma(h^*\rightarrow h\gamma)$. The
polarizabilities  are obtained from precision measurements of the $\gamma h
\rightarrow \gamma h$ $\gamma$-hadron Compton scattering. For the pion and kaon,
chiral perturbation theory ($\chi$PT) leads to precision predictions for the
polarizabilities (\cite{hols1,babu2}). Precision measurements of polarizabilities
therefore subject the $\chi$PT techniques of QCD to new and serious tests.

\subsection{Pion Polarizabilities via Primakoff Scattering}

\indent
~~~For the pion polarizability, $\gamma\pi$ scattering was measured (with large
uncertainties)  with 40 GeV pions (\cite{anti1}) via radiative pion scattering
(pion Bremsstrahlung) in the nuclear Coulomb field:
\begin{equation}
\label{eq:polariz}
\pi + Z \rightarrow \pi' + \gamma + Z'.
\end{equation}
In this measurement, the incident pion Compton scatters from a virtual photon in
the Coulomb field of a nucleus of atomic number Z; and the final state $\gamma$
and pion are detected in coincidence.  The radiative pion scattering reaction is
equivalent to  $\gamma$ + $\pi^{-}$ $\rightarrow$  $\gamma$ + $\pi^{-}$
scattering for laboratory $\gamma$'s of order 1 GeV incident on a target
$\pi^{-}$ at rest. It is an example of the well tested Primakoff formalism
(\cite{jens,ziel2}) that relates processes involving real photon interactions to
production cross sections involving the exchange of virtual photons.

In the 40 GeV radiative pion scattering experiments, it was shown experimentally
(\cite{anti1}) and theoretically (\cite{galp}) that the Coulomb amplitude clearly
dominates, and yields sharp peaks in t-distributions at very small squared four
momentum transfers (t) to the target nucleus t $\leq 6 \times 10^{-4}$
(GeV/c)$^{2}$. Backgrounds from strong  processes were low. The backgrounds are
expected to be lower at the higher energy (280 GeV) planned for the  CERN COMPASS
experiment.

{\bf All polarizabilities in this paper are expressed in  Gaussian units of
$10^{-43}$ cm$^3$}. The $\chi$PT prediction (\cite {hols1}) for the pion
polarizability is $\bar{\alpha}_{\pi}$ = 2.7. \cite {hols1} showed that meson
exchange via a pole diagram involving the a$_1$(1260) resonance provides the main
contribution ($\bar{\alpha}_{\pi}$ = 2.6) to the polarizability. \cite{xsb} assuming
a$_1$ dominance find $\bar{\alpha}_{\pi}$ = 1.8. For the kaon, the $\chi$PT
polarizability prediction (\cite {hols1}) is $\bar{\alpha}_{\pi}$ = 0.5. A more
extensive theoretical study of kaon polarizabilities was given recently (\cite {ev}).

\subsection{Pion Polarizabilities via Inelastic $\pi e$ Scattering}

Pion-electron elastic scattering $\pi e \rightarrow \pi' e'$ has been studied in
SELEX (\cite {selex}) at Fermilab with 600 GeV energy pion beams and electron
targets (atomic electrons in nuclear targets) to measure the low momentum part of
the pion form factor, and thereby the charge radius of the pion. Sigma-electron
scattering was also studied. These reactions were studied via a trigger that
required two negative charged particles in the final state, both over 25 GeV.
Various methods of particle identification are utilized to assure that one of the
final state particles is an electron. The electron signature together with
energy/momentum balance assure that the incident hadron interacts with an
electron, and not with the target nucleus.

With the same trigger, one may also study the virtual Compton scattering (VCS)
process $\pi e \rightarrow \pi' e' \gamma$ related to the generalized pion
polarizabilities $\bar{\alpha}_{\pi}$(k) and $\bar{\beta}_{\pi}$(k), which depend
on momentum transfer (k) to the electron (\cite {dd,guic}). In the limit of zero
momentum transfer, these reduce to the usual Compton polarizabilities. For the VCS
reaction, the Bethe-Heitler (BH) amplitude ($\gamma$ from initial or final state
electron, not from the pion Compton amplitude) dominates over the Compton amplitude.
For VCS on the pion,  the Compton amplitude should be relatively more enhanced
compared to BH for events in which the angle between $\gamma$ and electron is large.
The VCS process and planned VCS experiments (for the proton) at electron accelerators
have been discussed extensively in the present workshop. Theoretical calculations and
simulations are in progress (\cite {dd}) for pion VCS to understand the sensitivity
to the generalized polarizabilities for this reaction. Data from SELEX are being
analyzed for pion VCS. It is not clear at this stage of the analysis whether or not
the signal to background in SELEX will be sufficiently good to get quality data for
the VCS process.

\subsection{Radiative Transitions via Primakoff and Inelastic Electron Scattering}

In addition to polarizability measurements, COMPASS may also study radiative
transitions of incident mesons to higher excited states. COMPASS may obtain new data
(\cite {hadron1}) for radiative transitions leading from the pion to the $\rho$,
a$_1$(1260), and a$_2$(1320); and for the kaon to K$^*$. Searches for exotic mesons
(hybrids) are also possible in this way (\cite {hadron1}). The $\rho$ data is
obtained with a $\gamma\pi$ trigger, while most of the others require a particle
multiplicity trigger. Radiative transition widths are predicted by vector dominance
and quark models. Independent and higher precision data for these and higher
resonances would be valuable in order to allow a more meaningful comparison with
theoretical predictions. For example, the $\rho \rightarrow \pi \gamma$ width
measurements (\cite{jens}) range from 60 to 81 keV; and the a$_1$(1260) $\rightarrow
\pi \gamma$ width measurement (\cite{ziel}) is $0.64 \pm 0.25 $ MeV. FNAL SELEX data
are being analyzed now to identify the $\pi e \rightarrow \rho e'$ inelastic electron
scattering reaction. Clean data for this reaction would allow a determination of the
$\rho \rightarrow \pi \gamma$ radiative width from a measure of the transition form
factor (near zero momentum transfer).

\subsection{Chiral Anomaly}

Another interesting meson radiative transition involves the chiral anomaly term of
the effective Chiral Lagrangian. COMPASS may study the Chiral Axial Anomaly with
50-280 GeV pion beams with the same $\gamma\pi$ trigger as needed for polarizability.
The abnormal intrinsic parity (chiral anomaly) component of the effective Chiral
Lagrangian predicts (\cite{hols2,ca}) $F_{3\pi}$ = 9.7 GeV$^{-3},~ O(p^4),$ for the
$\gamma \rightarrow  3 \pi$ F$_{3\pi}$ amplitude at threshold. F$_{3\pi}$ was
measured (\cite{anti1}) with 40 GeV pions. They studied pion production by a pion in
the nuclear Coulomb field near threshold via the Primakoff reaction: \begin{equation}
 \pi^- + Z \rightarrow {\pi^-}' + \pi^0 + Z'. \label{eq:anomaly} \end{equation} This
reaction is equivalent to $\pi^- + \gamma \rightarrow {\pi^-}' + \pi^0$, a radiative
transition to a low mass two-pion system. The cross section for this Primakoff
reaction is proportional to $F_{3\pi}^2$. Low-t events were selected in the analysis
in order to isolate the Primakoff process. Diffractive production of the two pion
final state via Pomeron exchange is blocked by G-parity conservation. The low
statistics ($\sim$ 200) experiment (\cite{anti1}) reported
F$_{3\pi}$=$12.9\pm0.9$(stat)$\pm0.5$(sys)~GeV$^{-3}$, which differs from the
O(p$^4$) expectation. More precise data are needed for this amplitude. The expected
number of near threshold two-pion events in COMPASS is several orders of magnitude
larger than in all previous experiments (\cite{ca}).

With a 600 GeV pion beam and a target electron, one may also study $e \pi \rightarrow
e' \pi' \pi^0$ events, where the two $\gamma$'s detected in the $\gamma$ calorimeter
have a $\pi^0$ invariant mass, and the $\pi\pi^0$ system has invariant mass lower
than the $\rho$. Such inelastic data provides a means complementary to the Primakoff
scattering to determine the chiral anomaly transition form factor (\cite {hols2}) and
amplitude $F_{3\pi}$. Analysis of SELEX data in search of this reaction is in
progress.

\section{Pion Polarizabilities}

For the $\gamma\pi$ interaction at low energy, $\chi$PT provides a rigorous way to
make predictions via a Chiral Lagrangian written in terms of renormalized coupling
constants L$^r_i$ (\cite{gass1}). With a perturbative expansion of the effective
Lagrangian, the method establishes relationships between different processes in terms
of the L$^r_i$. For example, the radiative pion beta decay and electric pion
polarizability are expressed as (\cite{hols1}):

\begin{equation}
F_A/F_V = 32\pi^2(L^r_9+L^r_{10});~ \bar{\alpha}_{\pi} = 
\frac{4\alpha_f}{m_{\pi}F^{2}_{\pi}}(L^r_9+L^r_{10});
\label{eq:fafv}
\end{equation}
where F$_\pi$ is the pion decay constant, F$_A$ and F$_V$  are the axial vector and
vector coupling constants in the decay, and $\alpha_f$ is the fine structure
constant. The experimental ratio F$_A$/F$_V$  = 0.45 $\pm$ 0.06, leads to
$\bar{\alpha}_{\pi}$ = -$\bar{\beta}_{\pi}$ = 2.7 $\pm$ 0.4, where the error shown is
due to the uncertainty in the F$_A$/F$_V$ measurement (\cite{hols1,babu2}).

The pion polarizabilities deduced by \cite{anti1} in their low statistics
experiment ($\sim$ 7000 events) were $\bar{\alpha}_{\pi} = -\bar{\beta}_{\pi} =
6.8 \pm 1.4 \pm 1.2$. It was assumed in the analysis that $\bar{\alpha}_{\pi} +
\bar{\beta}_{\pi} = 0$, as expected theoretically (\cite {hols1}).  The deduced
polarizability value, not counting the large error bars, is some three times
larger than the $\chi$PT prediction. {\bf The available polarizability results
have large uncertainties. There is a clear need for new and improved radiative
pion scattering data.}

\section{Experimental Requirements}

\begin{figure}[tbc]
\centerline{\rotate[r]{\epsfig{file=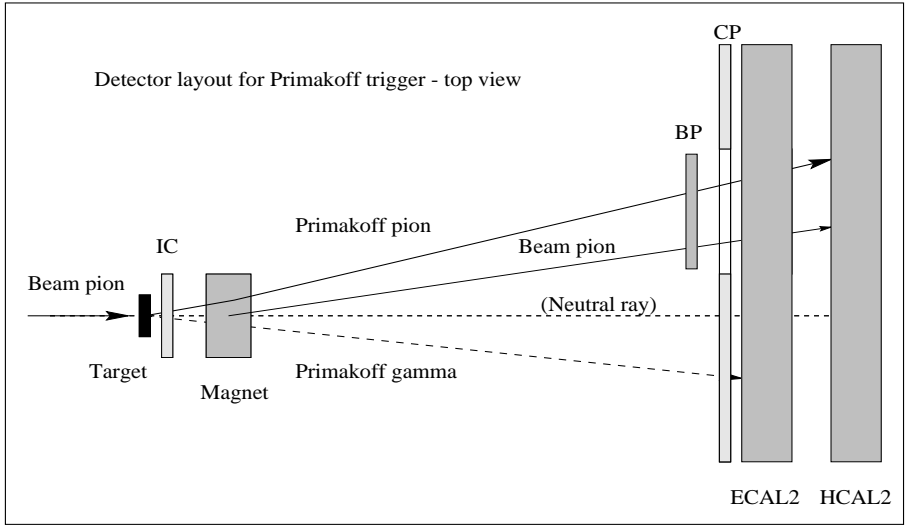,
width=7cm,height=12cm}}}
\caption{Detector layout for Primakoff trigger.}
\label{fig:trigger}
\end{figure}

We considered the beam, detector, and trigger requirements for polarizability,
chiral anomaly, and hybrid meson  studies in the CERN COMPASS experiment (\cite
{hadron1}). We begin for illustration with pion polarizability measurements via
the reaction $\pi^{-}$ + Z $\rightarrow$  $\pi^{-'}$  + $\gamma$ + Z$'$ with a
300 GeV pion beam. The beam energy is chosen to be maximal, since that pushes the
energy spectrum of final state $\gamma$'s and  $\pi^0$'s to be highest, and thereby
the detection acceptance for $\pi^0$'s for a given size ECAL2 electromagnetic
calorimeter will be maximal. Higher beam energy also gives a higher acceptance for
Primakoff production of high mass mesons. Fig.~\ref{fig:trigger} shows the detector
layout for this experiment. Proceeding downstream, we consider the scintillation
detectors IC (Interaction Counter), BP (Beam/Primakoff fiducial detector), CP
(Charged Particle), and the $\gamma$/hadron calorimeters ECAL2/HCAL2. The function of
these detectors in the trigger will be described below.

\subsection{Monte Carlo Simulations}

\begin{figure}[tbc]
\centerline{\epsfig{file=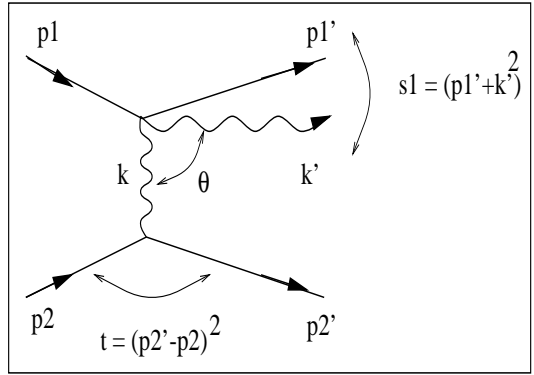,width=7cm,
height=5cm}}
\caption{The Primakoff $\gamma$-hadron Compton process and kinematic variables
(4-momenta): p1, p1$^\prime$ = for initial/final hadron, p2, p2$^\prime$ = for 
initial/final target, k, k$^\prime$ = for initial/final gamma, and $\theta$ 
the scattering angle of the $\gamma$ in the alab frame.}
\label{fig:diagram}
\end{figure}

We carried out Monte Carlo simulations with the code POLARIS , an event generator for
polarizability studies; and ANOMALY, for chiral anomaly studies (\cite{pol}). In this
report, we show only the POLARIS results. POLARIS produces events of the type
Eq.~\ref{eq:polariz}, based on the theoretical Primakoff $\gamma\pi$ Compton
scattering cross section. The four-momentum of each particle is p1, p2, p1$^\prime$,
p2$^\prime$, k, k$^\prime$, respectively, as shown in Fig.~\ref{fig:diagram}. In the
one-photon exchange domain, this reaction is equivalent to $\gamma + \pi  \rightarrow
 \gamma^\prime + \pi^\prime$, and the four-momentum of the incident virtual photon is
k = p2$-$p2$^\prime$. We have therefore t=k$^2$ with t the square of the
four-momentum transfer to the nucleus, F(t) the nuclear form factor (essentially
unity at small t, $\sqrt{\rm{s}}$ the mass of the $\gamma\pi$ final state, and t$_0$
the minimum value of t to produce a mass $\sqrt{\rm{s}}$. The momentum modulus
$|\vec{k}|$ (essentially equal to p$_T$) of the virtual photon is in the transverse
direction, and is equal and opposite to the momentum p$_T$ transferred to the target
nucleus. For the generated events, the pion and $\gamma$ laboratory variables may be
given gaussian spreads to simulate measurement errors, and acceptance cuts may be
used (optional). Finally, the simulated events are taken to be these "measured"
values. The pion polarizability is extracted via a fit of the theoretical cross
section to the scattered $\gamma$ angular distribution in the projectile (alab) rest
frame. The total Primakoff cross section is computed by integrating numerically the
differential cross section $\sigma(s,t,\theta)$ of Eq.~\ref{eq:Primakoff_1} below for
the Primakoff Compton process.

The code ANOMALY produces events with the topology of Eq.~\ref{eq:anomaly}, following
the techniques of POLARIS.

\subsection{Primakoff $\gamma\pi$ Compton Event Generator}

We describe the event generator for the radiative scattering of the pion in the
Coulomb field of a nucleus (\cite{pol}). In the pion alab frame, the nuclear Coulomb
field effectively provides a virtual photon beam incident on a pion target at rest.
We have for the variable  t=k$^2 \equiv$ M$^2$, where k is the 4-momentum transferred
to the nucleus, and M is the virtual photon mass. Since
t=2M$_Z$[M$_Z$-E(Z',lab)]$<$0, the virtual photon mass is imaginary. To approximate
real pion Compton scattering, the virtual photon should be taken to be almost real.
For small t, the electromagnetic contribution to the scattering amplitude is large
compared to meson and Pomeron exchange contributions.

The Primakoff differential cross section of the process of Eq.~\ref{eq:polariz} in
the alab frame may be expressed as (\cite{starkov}):
\begin{equation}
\label{eq:Primakoff_1}
\frac
{{d}^3{\sigma}}
{{dt}{d}{\omega}{d\cos{\theta}}}
=
\frac
{\alpha_{f}{Z}^2}
{\pi\omega}
\cdot
\frac
{t-t_{0
}}
{t^2}
\cdot
\frac
{{d}\sigma_{\gamma\pi}{(}\omega,\theta{)}}
{{d}{\cos}{\theta}},
\end{equation}
where the $\gamma\pi$ cross section is given by:
\begin{equation}
\label{eq:Primakoff_2}
\frac
{{d}\sigma_{\gamma\pi}{(}\omega,\theta{)}}
{{d}{\cos}{\theta}}
=
\frac
{{2}{\pi}{\alpha_{f}}^2}
{{m}_{\pi}^2}
\cdot
\{
{F}_{\gamma\pi}^{pt}{(}{\theta}{)}
+
\frac
{{m_{\pi}}{\omega}^2}
{\alpha_{f}}
\cdot
\frac
{\bar{\alpha}_{\pi}{(}1+{\cos}^{2}{\theta}{)}+2\bar{\beta}_{\pi}{\cos\theta}}
{{(}{1+\frac{\omega}{m_{\pi}}{(1-\cos{\theta})}}{)}^3}
\}.
\end{equation}
Here, t$_0$=$(m_{\pi}\omega/p_{b})^2$, with $p_{b}$ the incident pion beam
momentum in the laboratory, $\theta$ the scattering angle of the real photon
relative to the incident virtual photon direction in the alab frame, $\omega$ the
energy of the virtual photon in the alab frame, $Z$ the nuclear charge, $m_\pi$
the pion mass, $\alpha_{f}$ the fine structure constant, and $\bar{\alpha_\pi}$,
$\bar{\beta_\pi}$ the pion electric and magnetic polarizabilities. The energy of
the incident virtual photon in the alab (pion rest) frame is:
\begin{equation}
\omega \sim  (s - {m_{\pi}}^2)/2m_{\pi}.
\label{eq:omega}
\end{equation}
The function ${F}_{\gamma\pi}^{pt}{(}{\theta}{)}$ describing the Thomson cross
section for $\gamma$ scattering from a point pion is given by:
\begin{equation}
\label{eq:Primakoff_3}
{F}_{\gamma\pi}^{pt}{(}{\theta}{)}=
\frac{1}{2}\cdot
\frac
{1+{\cos}^{2}{\theta}}
{{(}{1+\frac{\omega}{m_{\pi}}{(1-\cos{\theta})}}{)}^2}
{.}
\end{equation}

From Eq.~\ref{eq:Primakoff_2}, the cross section depends on
$(\bar{\alpha}_{\pi}+\bar{\beta}_{\pi})$ at small $\theta$, and on
$(\bar{\alpha}_{\pi}-\bar{\beta}_{\pi})$ at large $\theta$. A precise fit of the
theoretical cross section (Eq.~\ref{eq:Primakoff_1}-\ref{eq:Primakoff_3}) to the
measured angular distribution of scattered $\gamma$'s, allows one to extract the
pion electric and magnetic polarizabilities. Fits will be done for different
regions of $\omega$ for better understanding of the systematic uncertainties. We
will carry out analyses with and without the dispersion sum rule constraint
(\cite {hols1}) that $\bar{\alpha}_{\pi}+\bar{\beta}_{\pi}\approx0.4$. We can
achieve a significantly smaller uncertainty for the polarizability by including
this constraint in the fits.

The event generator produces events in the alab frame, characterized by the
kinematical variables t, $\omega$ and $\cos\theta$, and distributed with the 
probability, of the theoretical Compton Primakoff cross section
(Eq.~\ref{eq:Primakoff_1}-\ref{eq:Primakoff_3}). Then, the $\gamma\pi$ scattering
kinematics are calculated. The virtual photon incident along the recoil direction
$\vec{k}/|\vec{k}|$, is scattered on the pion "target", and emerges as a real photon
with energy/momentum $\omega^{\prime}$/$|\vec{k}^{\prime}|$ at an angle $\theta$:
\begin{equation}
\label{eq:Compton}
{\large
\omega^{\prime}=\frac
{\omega{(}{1}+\frac
{\omega^2{-}{|\vec{k}|}^2}{{2}{m}_{\pi}{\omega}}
{)}}
{
{1}{+}
\frac{\omega}{{m}_{\pi}}
{(}{1-}
\frac{|\vec{k}|}{\omega}
\cos\theta
{)}
}
}
\end{equation}

The photon azimuthal angle around the recoil direction is randomly generated
using a uniform distribution. The four-vector components of all reaction
participants (pion, photon and recoil nucleus) are then calculated in the alab
frame. The azimuthal angle of the recoil nucleus is also randomly generated by a
uniform distribution. Finally, the reaction kinematics are transformed to the lab
frame by a Lorentz boost.

\begin{figure}[tbc]
\centerline{\epsfig{file=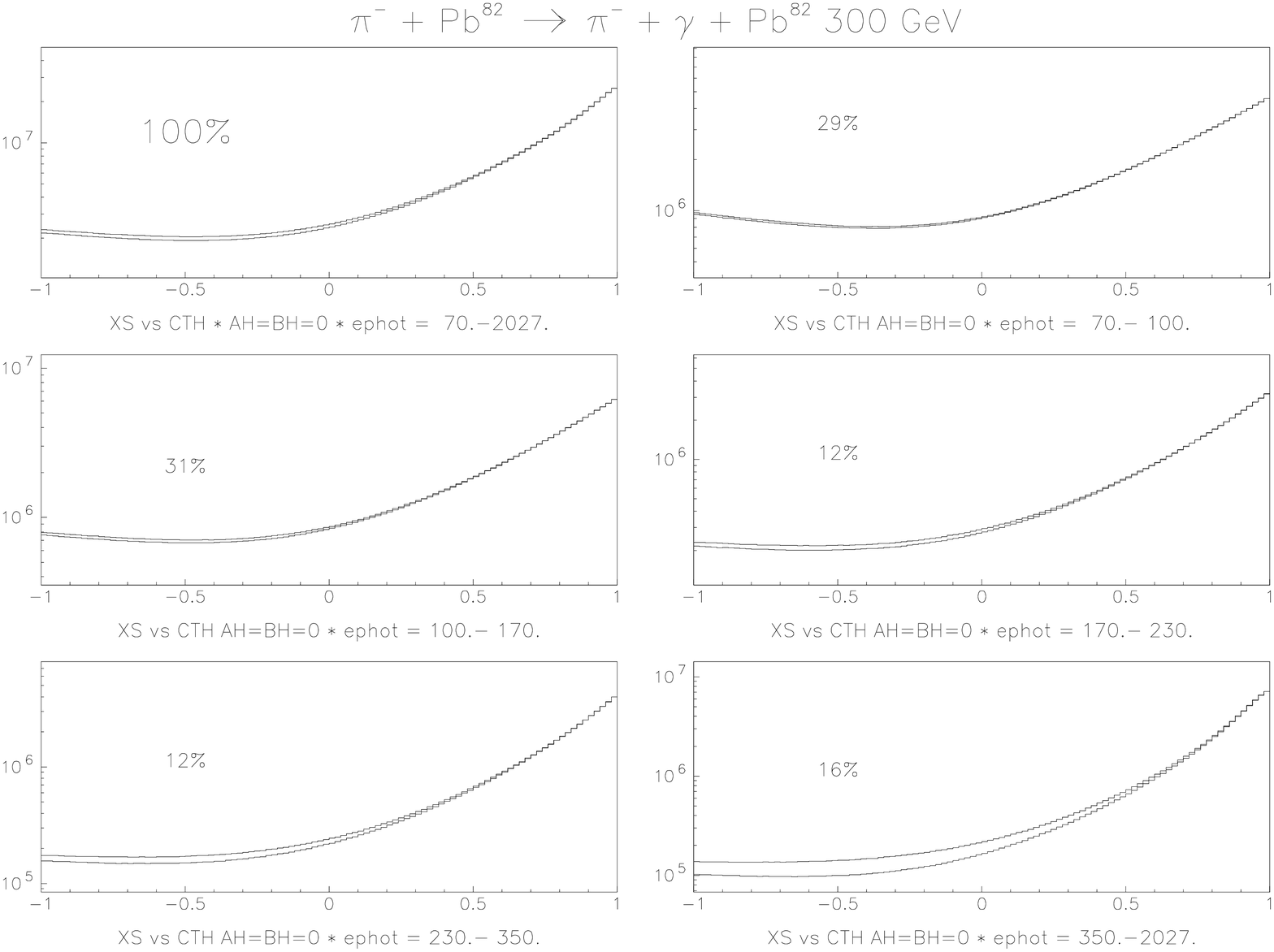,
width=12cm,height=7cm}}
\caption{The dependence of the theoretical angular distributions on
polarizability for different regions of $\gamma$ energy $\omega$ 
(given in MeV), function of $\cos(\theta)$ in the alab frame.
The lower curve corresponds to $\bar{\alpha}$=7,
$\bar{\beta}=-$6; while the upper curve corresponds
to zero polarizabilities. The percentage shows the statistics fraction
in each $\omega$ region.}
\label{fig:cth_sens_1}
\end{figure}

\begin{figure}[tbc]
\centerline{\epsfig{file=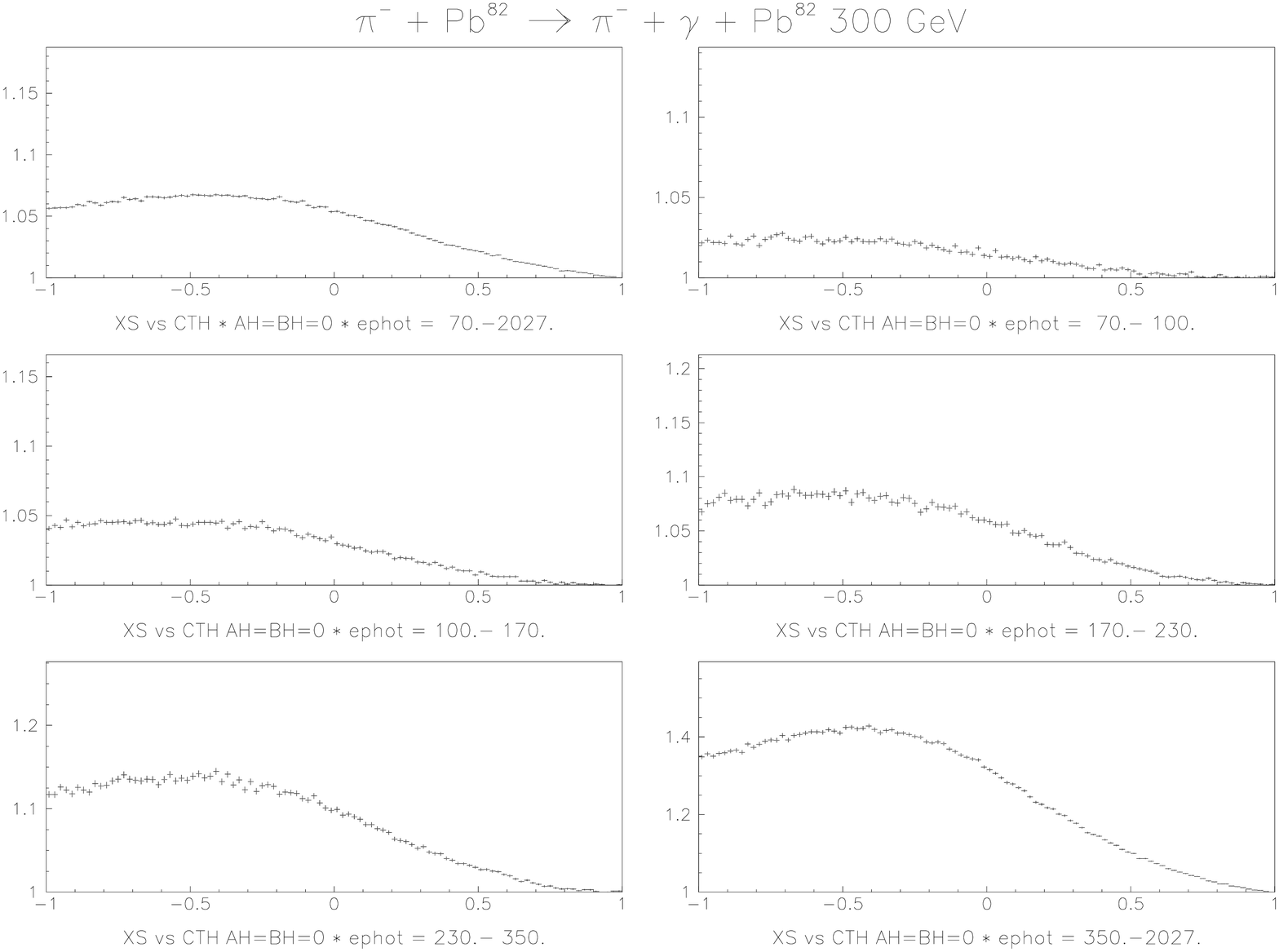,
width=12cm,height=7cm}}
\caption{Ratio of the theoretical angular distributions for different
regions of $\gamma$ energy $\omega$ (given in MeV), as a function of
$\cos(\theta)$ in the alab frame, for
the case of zero polariabilities (Thomson term only),
relative to the case in which
$\bar{\alpha}$ =7, $\bar{\beta}=-$6.
The contribution of the polarizability to the
cross section is larger at back alab angles, and increases with
increasing $\omega$.}
\label{fig:cth_sens_3}
\end{figure}

For the measurement of the pion polarizabilities, one must fit the theoretical cross
section (\ref{eq:Primakoff_1}-\ref{eq:Primakoff_3}) to measured distributions, after
correcting for acceptances. The sensitivity to the polarizability increases with
increasing $\omega$ energy and at back angles. A convenient method is to use the
$\cos\theta$ distribution integrated over t and $\omega$, since this shows clearly
the sensitivity to the polarizability. The sensitivities of the theoretical angular
distributions to the polarizabilities  (i.e., for $\bar{\alpha}$=$\bar{\beta}$=0 and
for $\bar{\alpha}$=7, $\bar{\beta}$=-6) for different regions of $\omega$ are given
in Fig.~\ref{fig:cth_sens_1}-\ref{fig:cth_sens_3}. The per cent statistics for
different $\omega$ regions are shown in Fig.~\ref{fig:cth_sens_1}.

\subsection{Design of the Primakoff Trigger\label{sec:how}}

The small Primakoff cross section and the high statistics required for extracting
polarizabilities require a data run at high beam intensities and with good
acceptance. This sets the main requirements for the trigger system: (1) to act as a
"beam killer" to suppress the high rate background associated with non-interacting
beam pions, (2) to avoid cutting the   acceptance at the important $\gamma$ back
angles in the alab frame, where the hadron polarizability measurement is most
sensitive, (3) to cope with background in the $\gamma$ calorimeter from low energy
$\gamma$'s or delta electrons.

COMPASS plans to construct a Primakoff trigger that incorporates a veto of the
non-interacting beam in a window on the hadron energy in HCAL2, and which includes a
coincidence of the scattered pion with a $\gamma$ measured in the ECAL2 calorimeter.
We studied the feasibility of such a trigger, via simulations carried out at 300 GeV
(\cite {hadron1}).

\begin{figure}[tbc]
\centerline{\epsfig{file=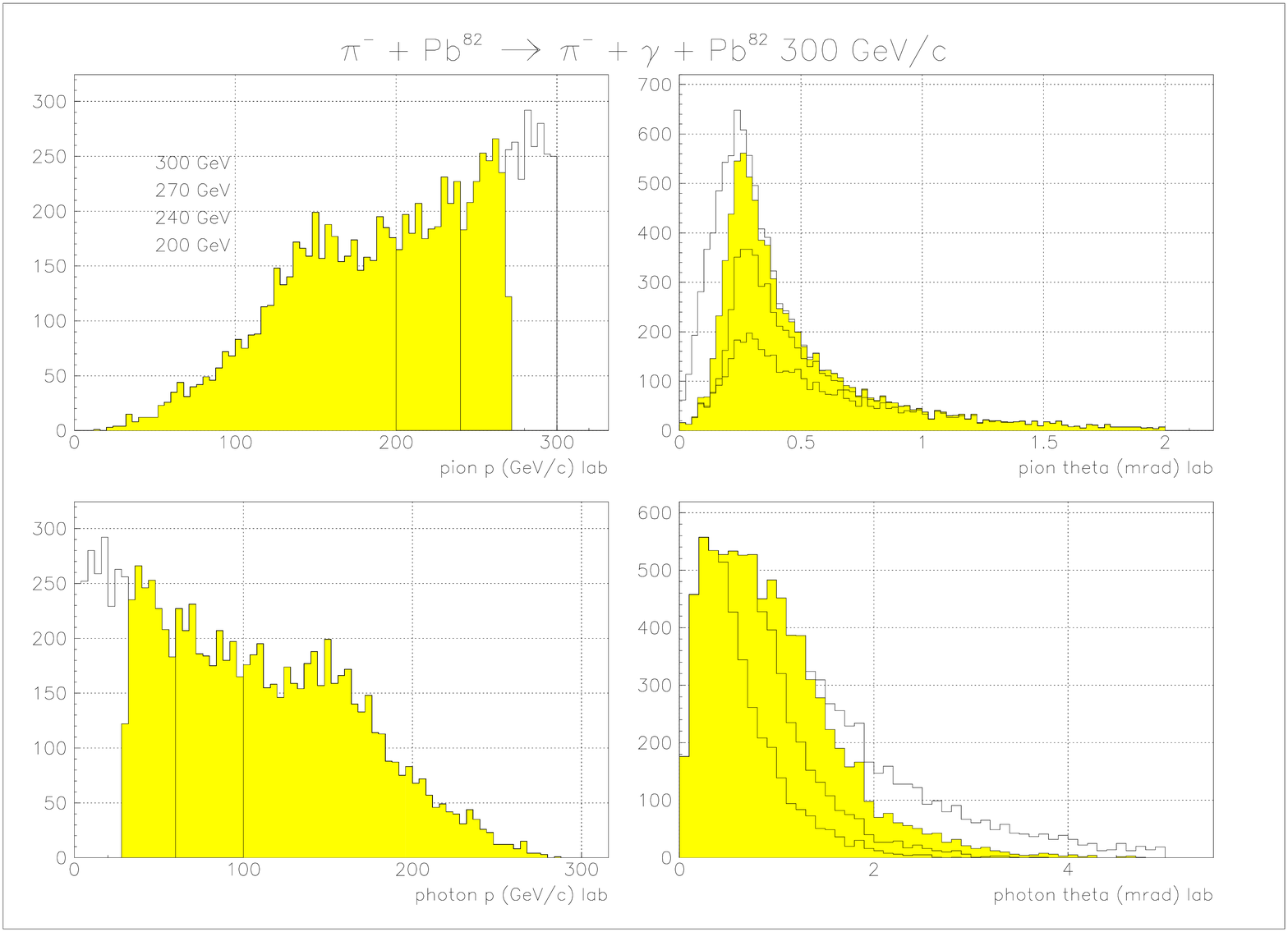,width=12cm,heig
ht=7cm}}
\caption{MC simulation showing the kinematics of the
$\gamma\pi\rightarrow\gamma\pi$ reaction, in terms of the $\pi$ and $\gamma$
momenta and angles. The overlayed spectra correspond to different trigger cuts
on the final state $\pi$ momentum.}
\label{fig:kinematics}
\end{figure}

For the reaction given in Eq. ~\ref{eq:polariz}, the laboratory outgoing
$\gamma$'s are emitted within an angular cone of within 5 mrad, and the
corresponding outgoing $\pi$'s are emitted within  2 mrad. Most events have
$\gamma$ energies between $0-280$ GeV, and $\pi$ energies between $20-300$ GeV.
The kinematics are shown in Fig.~\ref{fig:kinematics}. The recoil nucleus of mass
M for a Primakoff reaction has negligible recoil energy (T$_r\approx$t/2 M),
with roughly 99\% of the events having  recoil kinetic energies less than 30 keV.
The corresponding final state $\pi$ and $\gamma$ carry all the four momentum of
the beam pion. Momentum and energy conservation may be used at the analysis stage
for background suppression.

Our MC shows that we lose very little polarizability information by applying an
"energy cut" trigger condition that rejects events with the outgoing pion energy
having more than 240 GeV. Corespondingly, the final state $\gamma$ has less than
60 GeV. The 240 GeV cut value was devised to act as a beam killer, as discussed
in more detail below. The 60 GeV cut will also be very effective in reducing the
$\gamma$ detector (ECAL2) trigger rate, since a large part of the background
$\gamma$ rate is below 60 GeV.

\begin{figure}[tbc]
\centerline{\epsfig{file=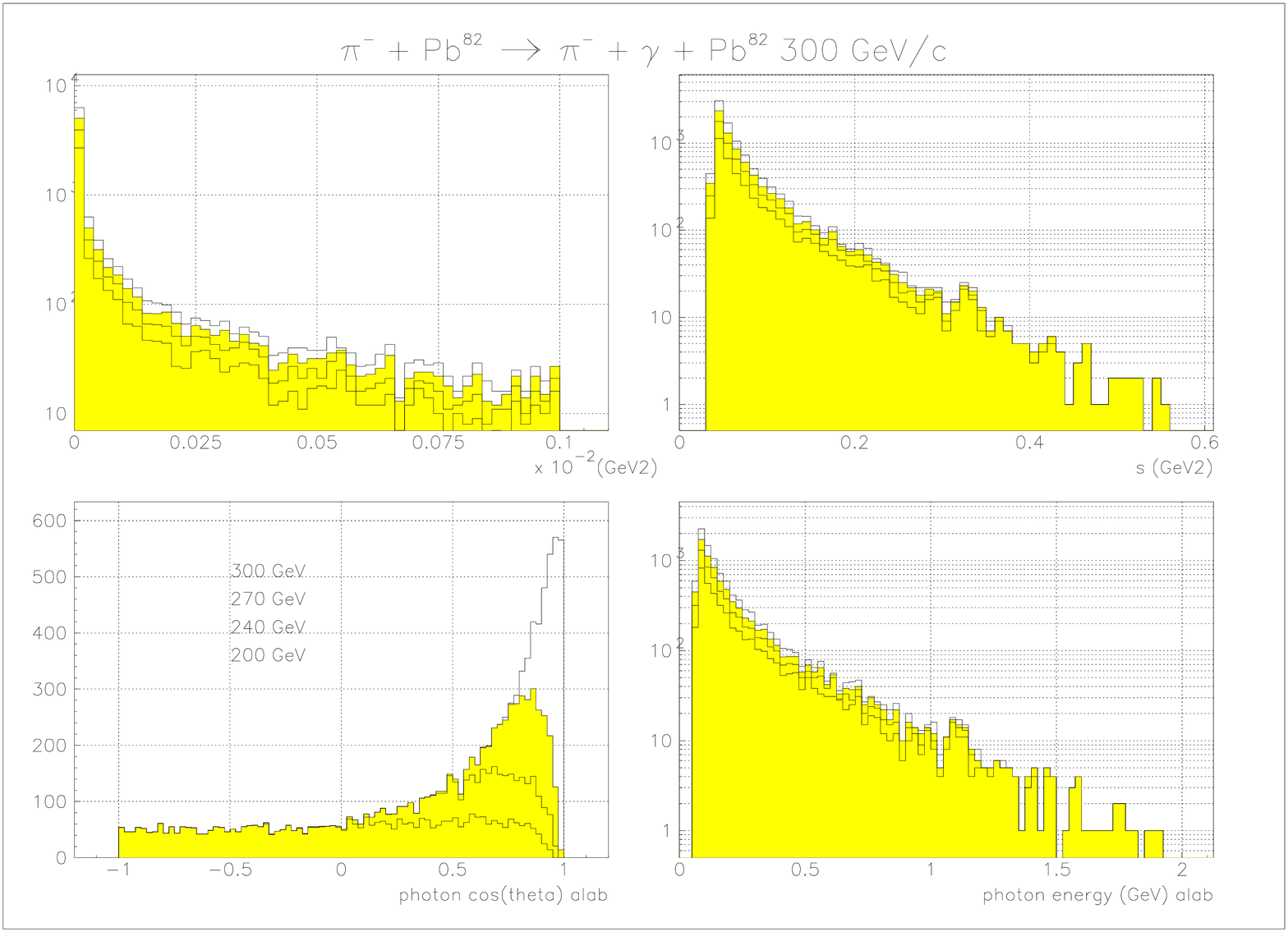,width=12cm,
height=7cm}}
\caption{MC simulation showing the acceptance of the
$\gamma\pi\rightarrow\gamma\pi$ reaction in terms of the invariant four
momentum transfer t to the target,the squared invariant energy s of the final
state $\gamma\pi$, the angular distribution versus $\cos(\theta)$ with
$\theta$ the $\gamma$ scattering angle in the alab frame, and the virtual photon
energy $\omega$ 
in the alab frame. The overlayed spectra correspond to different cuts
on the final state $\pi$ momentum.}
\label{fig:acceptance}
\end{figure}

The polarizability insensitivity to these cuts results from the fact that the
most forward (in alab frame) Compton scattering angles have the lowest laboratory
$\gamma$ energies and largest laboratory angles. In addition, the cross section in
this forward alab angle range is much less sensitive to the polarizabilities. This is
seen from Eq. \ref{eq:Primakoff_2}, since with
$\bar{\alpha}_{\pi}+\bar{\beta}_{\pi}\approx 1$ used in our MC, the polarizability
component is small at forward compared to  back angles. The acceptance is reduced by
the energy cut for the forward alab angles (shown in Fig.~\ref{fig:acceptance}), but
is unaffected at the important alab back angles. Summarizing, the pion and $\gamma$
energy constraints at the trigger level fulfill the "beam killer" requirement and at
the same time remove backgrounds coming from low energy $\gamma$'s,   delta
electrons, and e$^+$e$^-$ pairs incident on ECAL2, etc.

\subsection{Beam Requirements}

In COMPASS, two beam Cherenkov detectors (CEDARS) far upstream of the target provide
$\pi/K/p$ particle identification (PID). The incoming hadron momentum is measured in
the beam spectrometer. Before and after the target, charged particles are tracked by
high resolution silicon strip tracking detectors. The measurement of both initial and
final state momenta provides constraints to identify the reaction. The final state
hadron and $\gamma$ momenta are measured downstream in the magnetic spectrometer and
in the $\gamma$ calorimeter, respectively.  These measurements allow a precise
determination of the small p$_T$ kick to the target nucleus, the main signature of
the Primakoff process, and the means to separate Primakoff from diffractive
scattering events.

We can get quality statistics for the pion  study with high beam intensities at the
CERN SPS. Some of the detectors needed for this study (such as HCAL2 modules with a
signal duration of about 50 nsec) must accept the full beam intensity, and cannot
tolerate beam intensities higher than 5 MHz. We will take data with different beam
energies and targets, with both positive and negative beams, as part of efforts to
control systematic errors.

\subsection{Target and Target Detectors\label{sec:target}}

The main Primakoff target will be Pb which has approximately a 1.2 mb Compton
scattering (polarizability) cross section and total inelastic cross section of 1.8
barn. For a  COMPASS pion beam rate of 5 MHz during the 2.5 second beam spill (every
19 seconds), and a 1\% interaction Pb target, we therefore expect approximately 80
events per spill ($80 \approx 1.2/1.8  \times 10^{-3} \times 10^{-2} \times 5. \times
10^{6} \times 2.5$) from the pion Primakoff effect. We also need Primakoff scattering
on nuclei with Z$<82$ to check the expected Z$^2$ cross section dependence.

The target is followed by two scintillation interaction counters (IC) with a
triggering condition of 1 minimum ionizing particle (mip) each (see
Fig.~\ref{fig:trigger}).  We use Si tracking detectors before and immediately after
the targets. We veto target break-up events by selecting  1 mip in the IC counters
after the targets, and by selecting low-t events in the off-line analysis.

\subsection{The Magnetic Spectrometer and the t-Resolution}

We need good momentum resolution for the incident and final state pions and
$\gamma$'s. In this way, the important four momentum t resolution can be
kept as good as possible. A final state $\pi^-$ at 200 GeV/c can be momentum
analyzed to 2 GeV/c resolution, with better resolution at lower momenta.

The angular resolution for the final state $\pi$ can be controlled by minimizing the
multiple scattering in the targets and detectors. With a lead target of 1\%
interaction length (2 g/cm$^2$,~30\% radiation length), multiple Coulomb scattering
(MCS) of the beam and outgoing pion in the target gives an rms angular resolution of
order 40 $\mu$rad. The intrinsic silicon tracking detector angular resolution is
significantly better than the MCS contribution to the angular resolution. We estimate
the target contribution to the resolution of the transverse momentum p$_T$ by
considering the p$_T$ generated through MCS for a non-interacting straight-through
beam pion of 200 GeV. The p$_T$ given to such a beam pion (with no Compton
scattering) is then $p_T = p \times \Delta{\theta} = 200 \times 40 \times 10^{-6} =
8$ MeV, which corresponds to $t~=~p_T^2~=~0.6~\times~10^{-4}$ GeV$^2$. Including other
effects (\cite {hadron1}), we aim for a p$_T$ resolution less than 15 MeV.
corresponding to $\Delta t$ better than $\approx$ 2.5 $\times 10^{-4}$ GeV$^2$.

This resolution will allow an effective t-cut to minimize contributions to the data
from diffractive processes. The goal is achievable, based on the t distributions
measured at a 200 GeV low statistics, high resolution experiment for $\pi^-
\rightarrow \pi^- \pi^0$ (\cite{jens}) and $\pi^- \rightarrow \pi^- \gamma$
(\cite{ziel2}) Primakoff scattering at 200 GeV at FNAL. The t distribution of the
$\pi^- \rightarrow \pi^- \gamma$ data agrees well with the Primakoff formalism out to
t~=~$10^{-3}$ GeV$^2$, which indicates that the data are indeed dominated by Coulomb
production. Minimum material (radiation and interaction lengths) in COMPASS will also
give a higher acceptance, since that allows $\gamma$'s to arrive at ECAL2 with minimum
interaction losses, and minimum $e^+e^-$ backgrounds.

\subsection{The $\gamma$ Calorimeter ECAL2}

COMPASS will be able to measure a final state 200 GeV $\gamma$ to $\pm$2 GeV, with a
position resolution of 1.5 mm, in the second $\gamma$ calorimeter ECAL2. This
$\gamma$ detector is equipped with 3.8 by 3.8 cm$^2$ GAMS lead glass blocks to make a
total active area of order 1.5 m diameter. An exotic hybrid meson study ($\pi\eta$
detection) that will run simultaneously with the polarizability/anomaly study fixes
the area of ECAL2 (\cite {hadron1}). The area needed for the polarizability
measurement is only 30$\times$30 cm$^2$.

The p$_T$ kicks of the COMPASS magnets are 0.45 GeV/c for SM1 (4 meters from target)
and 1.2 GeV/c for SM2 (16 meters from target). We require the highest conveniently
accessible effective p$_T$ kick for this physics. The fields of both magnets must
therefore be set $\it{additive}$ for maximum deflection of the beam from the zero
degree (neutral ray) line. ECAL2 should be at maximum distance from the target
($\approx$ 40 meters) to also maximize the distance between the zero degree line and
the deflected non-interacting beam position. ECAL2 has a hole through which the beam
passes and
hits HCAL2.
 
We need to attain at least 10 cm for the distance between the zero degree line and
the hole edge. This is so since the Primakoff $\gamma$'s are concentrated around the
zero degree line and a good $\gamma$ measurement requires clean signals from 9
blocks, centered on the hit block. The beam hole size and position must be optimized
to minimize the hadrons hitting ECAL2 blocks at the hole perimeter. We plan it to be
big enough (2 blocks V $\times$ 16 blocks H) to pass completely the non-interacting
beam, and to pass also the majority of Primakoff scattered pions. In that way, these
particles are measured well in the HCAL2 hadron calorimeter behind ECAL2. We are then
able to optimally fix the beam killer threshold cut.

From MC simulations, the number of Primakoff scattered pions below 40 GeV is less
than 0.3\%, so that 40 GeV pions are about the lowest energy of interest. We will
effectively set a $\pi^-$ acceptance energy window of 40 - 240 GeV, via a minimum
threshold of  60 GeV for the $\gamma$ energy deposited in ECAL2, and an HCAL2 veto
for energies above 240 GeV.

\subsection{The Hadron Calorimeter HCAL2}

We intend to use beam rates of order 5 MHz, where the rate limit is the maximum
allowed for good operation of the existing and tested 15 $\times$ 15 cm$^2$ Dubna
hadron calorimeter  modules. For the beam killer trigger purposes, we require a
mini-HCAL2 configured  as an array of 15 $\times$ 15 cm$^2$ blocks (2 $\times$ 2 or 
3 $\times$ 3) to catch non-interacting beam pions. As shown in
Fig.~\ref{fig:trigger}, we will actually use a larger HCAL2. But the energy sum for
trigger purposes would still be taken from the mini-HCAL2. The HCAL2 modules have
energy resolution of $\pm$15 GeV at 300 GeV. Together with the beam acceptance of
$\pm$ 13 GeV, we can achieve a 1-$\sigma$ identification of the beam via a detection
window of 300 $\pm$ 20 GeV. We can therefore set a 3-$\sigma$ discriminator veto
threshold at $300-3\times20=240$ GeV, to veto 99\% of the beam. We will reduce the
beam acceptance to 13 GeV rms or lower, by collimation. In Table~\ref{tab:rate}, we
estimate the ECAL2/HCAL2 effect on the Primakoff trigger. The mini-HCAL2 modules
analog signals will be electronically summed and discriminated to provide a veto
trigger signal for hadron energies above 240 GeV.

\subsection{The Primakoff Trigger\label{sec:trigger}}

We design the Primakoff trigger using three trigger levels. The final trigger
signals should be developed in the minimum time possible (to reduce dead time),
and within the 300-1000 nsec allowed in the COMPASS data acquisition. The T0
trigger is a fast logic signal defining the beam phase space, rate and purity at
the target, and is generated near the target about 20 nsec after beam passage. It
is produced via a logic relation between signals from an ensemble of beam
transmission and beam halo veto (hole) scintillators located before the target.
The T1 trigger exploits the essential feature of a Primakoff polarizability (and
chiral anomaly) event; namely, a coincidence between a $\gamma$ in ECAL2 and a
scattered Primakoff pion. The detectors are shown in Fig.~\ref{fig:trigger}.

The IC counter logical signal should correspond to an amplitude of one 1 mip.
BP (Beam or Primakoff) is a scintillator fiducial trigger scintillation detector
with dimensions of order 60 cm (in H) by 15 cm (in V), which is the size covered
by the non-interacting pion beam and the Primakoff scattered pions. BP helps form
the pion detection trigger; it is set to fire on a 1 mip window condition. The
vertical size of this BP detector is larger than the 10 cm needed for
polarizability. This is so in order to catch also the scattered $\pi^-$'s
associated with the $\pi^-\pi^0$ and $\pi^-\eta$ final states from the chiral
anomaly and hybrid meson triggers, since there is a larger angular spread of the
$\pi^-$'s from these channels (\cite {hadron1}).

CP is a charged particle veto scintillator array positioned at the front face of
ECAL2. It is designed with a hole slightly larger than the BP detector, in which
the BP detector above is positioned. It covers the front face of ECAL2. CP
protects ECAL2 from charged particles. Simulations in progress will help fix the
definitive sizes of BP, CP, and the ECAL2 beam hole.

The first level trigger T1 is defined as:
\begin {equation}
T1 = IC (1 ~\rm{mip}) \cdot BP (1 ~\rm{mip}) \cdot \overline{CP}
         \cdot ECAL2 (> 60 \rm{GeV}) \cdot \overline{HCAL2} (>240 \rm{GeV}).
\end{equation}

\noindent
The trigger is designed to accept only events in which one Primakoff scattered
pion hits and fires IC and BP, the $\gamma$ energy exceeds 60 GeV, and HCAL2 does not
measure more than 240 GeV. All of the non-interacting $\pi$ beam and most of the
Primakoff scattered $\pi$'s pass through the ECAL2 hole. These pions proceed to
HCAL2, where their energy is measured well. The ECAL2 low energy threshold is
important to suppress low energy backgrounds.

The task of T1 is to provide a fast gate signal to start digitization (for
example in the ADC-system of the calorimeter) about $\sim$ 300 ns after the beam
traverses the target (see Table~\ref{tab:rate}). A second level trigger T2 can be
constructed if a faster T1 or more rate reduction is needed. A faster T1 is
possible if the IC counter 1 mip trigger signal, which arrives the latest at the
coincidence module near ECAL2, will be transferred from T1 to T2. Further rate
reduction may be gained using additional trigger conditions at the T2 level.
Details have been given elsewhere (\cite {hadron1}).

\begin{table}
\begin{center}
\begin{tabular}[tbc]{|l|c|c|c|c|}
\hline
Signature & Amp. range & Timing (nsec) & Reduc. fact.& Rate (events/spill)  \\
\hline
Beam                    & $-$           & 0   & $-$ & 1.25~10$^7$  \\
IC (interaction counter)& 1 mip         & 1   & $-$ & 1.25~10$^7$  \\
BP (beam or Primakoff)  & 1 mip         & 200 & $-$ & 1.25~10$^7$  \\
CP (charged particles)  & $\geq1$ mip   & 200 &   5 & 2.5~$10^6$   \\
HCAL2 ($\pi$ energy)    & $<240$ GeV  & 260 & $-$ & $-$            \\
                        &               &     & 1000  & 2.5~$10^3$ \\
ECAL2 ($\gamma$ energy) & $>60$ GeV  & 260 & $-$ & $-$             \\
\hline
\end{tabular}
\end{center}
\caption{The Primakoff trigger conditions and estimation of timing
relative to the target crossing time, and trigger rate reduction. For HCAL2
and ECAL2, we consider coincidences and a common reduction factor.}
\label{tab:rate}
\end{table}

\subsection{Expected Trigger Rates  \label{sec:trig1}}

\indent
~~~The 
ECAL2 $\gamma$ signal above 60 GeV with $\overline{HCAL2}(>240~\rm{GeV})$ and
in coincidence with BP should reduce the trigger rate from the beam rate by an
estimated factor of 1000. The CP detector requirement should give at least
another factor of 5 rate reduction. In this way, one may expect to achieve a
trigger rate lower than the beam rate by a factor of 5000 (\cite{hadron1}). The
rate of this signal ($2.5 \times 10^3$ per spill in Table ~\ref{tab:rate}) will
be significantly lower than the maximum of $10^5$ per spill trigger rate
planned for the COMPASS data acquisition.

We need to study more precisely the background rates, and ways to reduce
backgrounds. For this purpose, we will use an event generator for pion-nucleus
interactions, embedded in the COMPASS apparatus. We will study what fraction of
the events generated pass our trigger conditions. The factor 1000 reduction above
is only a guess of what we expect from the $\gamma\pi$ coincidence condition. The
trigger conditions are summarized in Table~\ref{tab:rate}.

\subsection{Measurement Significance}

The experimental pion polarizability determination to date has large
uncertainties; and kaon polarizabilities have never been measured. We
will determine the $\gamma\pi$ and $\gamma$K Compton cross section in tha alab
frame versus $\omega$ and cos($\theta$). We consider now the uncertainties
achievable for the pion polarizabilities in the COMPASS experiment, based on
Monte Carlo simulations.

We estimated 80 events/spill from the pion Primakoff effect (see
Sec.~\ref{sec:target}), corresponding to $10^{7}$ events per month at 100\%
efficiency. We assume a trigger efficiency of 50\% (due to the energy cuts), an
accelerator operating efficiency of 50\%, and a tracking efficiency of 80\%. One may
then expect to observe as many as $2 \times 10^{6}$ Primakoff Compton events per
month of operation, following setup of COMPASS. Statistics of this order will allow
systematic studies, with fits carried out for different regions of $\omega$, Z$^2$,
etc.; and  polarizability determinations with statistical uncertainties of order 0.2.
For the kaon polarizability, due to the lower beam intensity, the statistics will be
roughly 50 times lower. A precision kaon polarizability measurement requires more
data taking time.

Comparing chiral anomaly to polarizability data, we expect roughly 300 times 
lower statistics, due to the 140 times lower cross section and the lower 
$\pi^0$ detection efficiency (\cite {ca}).

\section{Conclusions}

The beams at CERN invite hadron Compton scattering and radiative transition
studies for different particle types, such as $\pi^{+,-}$, $K^{+,-}$, $p$, $\bar
{p}$, and others. COMPASS will measure the $\gamma\pi$ and $\gamma$K Compton
scattering cross sections, thereby enabling determinations of the pion and Kaon
polarizabilities. COMPASS will also measure the formation and decay of the
a$_1$(1260) and other resonances, and  also the chiral anomaly amplitude
F$_{3\pi}$. The pion and Kaon  experiments will allow serious tests of $\chi$PT;
and of different available polarizability and radiative decay calculations in
QCD. We also described FNAL SELEX E781 attempts to  study related physics via the
interaction of 600 GeV pions with target electrons.

\section{Acknowledgments}

This research was supported by the U.S.-Israel Binational Science Foundation (BSF)
and the Israel Science Foundation founded by the Israel Academy of Sciences and
Humanities, Jerusalem, Israel. Special thanks are due to the M.P.I. Heidelberg
SELEX/COMPASS group, U. Dersch, F. Dropmann, I. Eschrich, H. Kruger, J. Pochodzalla,
B. Povh, J. Simon, and K. Vorwalter, for hospitality and collaboration during the
writing of this report. Thanks are due to M. Buenerd, D. Drechsel, T. Ferbel, L.
Frankfurt, A. Ocheraschvili, S. Paul, J. Russ, I. Savin, H.-W. Siebert, A. Singovsky,
and  T. Walcher for valuable discussions.

\end{document}